# Lateral length scales in exchange bias.

Igor V. Roshchin,[1] O. Petracic,[1,2] R. Morales,[1,3] Zhi-Pan Li,[1] Xavier Batlle,[1,4] and Ivan K. Schuller[1]

[1]Physics Department, University of California - San Diego, La Jolla, CA 92093-0319
[2]Angewandte Physik, Universität Duisburg-Essen, 47048 Duisburg, Germany
[3]Departamento de Fisica, Universidad de Oviedo, Oviedo 33007, Spain
[4]Departament de Física Fonamental, Universitat de Barcelona, 08028 Barcelona, Catalonia, Spain.

When a ferromagnet is in proximity with an antiferromagnet, *lateral* length scales such as the respective magnetic domain sizes drastically affect the exchange bias. Bilayers of $FeF_2$ and either Ni, Co or Fe are studied using SQUID and spatially resolved MOKE. When the antiferromagnetic domains are larger than or comparable to the ferromagnetic domains, a local, non-averaging exchange bias is observed. This gives rise to unusual and tunable magnetic hysteresis curves.

Understanding the relevant length scales that characterize a particular phenomenon or type of interaction in a material is one of the most important issues in physics. This becomes less obvious and more intriguing when two dissimilar materials are in contact. In such cases, the proximity effect is often observed, where one material modifies the properties of the other. Because of the finite extent of electron wave functions, the proximity effect is typically described as the spatial variation of an order parameter *across* the interface. For example, the proximity effect that occurs in superconductor-normal bilayers, is characterized by the decay of the superconducting

2order parameter into the normal material with a length scale referred to as the coherence length. Interestingly, no correlation between characteristic length scales of physical quantities changing in the plane *parallel* to the interface has ever been claimed to play a role in the proximity effect. For instance, in case of the superconducting proximity effect between a conventional, *s*-wave superconductor and a normal metal, the order parameter of the superconductor is assumed spatially uniform in the plane parallel to the interface; thus, lateral length scales play no role.

In this work, we demonstrate for the first time that the *relation* between the *lateral* characteristic length scales on the two sides of the interface is important for exchange bias. Exchange bias (EB) is a proximity effect between a ferromagnet (F) and an antiferromagnet (AF) in intimate contact with each other [1]. Usually, EB is described as an additional unidirectional anisotropy induced by the AF into the F via exchange coupling at the interface. This produces a single magnetic hysteresis loop shifted along the magnetic field axis, below the AF ordering (Néel) temperature, $T_N$. The magnitude of this shift is defined as the exchange bias field, $H_e$, and can be positive or negative. Less obvious and usually more difficult to observe, is the effect of the F on the properties of the AF [2]. Both the F and AF can form magnetic domains, which leads to spatial variation of the magnetization in the F and the staggered magnetization in the AF in the plane parallel to the interface. Due to the mutual influence between the F and AF, the interfacial spin configuration may deviate significantly from that of the bulk AF and F. In this report, the AF domains in the exchange-biased systems refer to the area that induces the same unidirectional anisotropy.

We demonstrate that when the AF domains are larger than or comparable to the F domains, local, non-averaging EB is observed. Such a sample can split magnetically into two subsystems with the $H_e$ of the same magnitude but opposite sign. This state is achieved either by zero-field cooling a partially demagnetized sample or by cooling the sample in a properly chosen constant applied magnetic field. In either case, the two magnetic subsystems behave independently of each other, and no averaging of EB occurs. This manifests itself clearly as a double hysteresis loop. In addition, the local EB sign can be manipulated by either creating domains in the F during the demagnetization process or by varying the field cooling procedure. These results reveal new physics of heterogeneous magnetic structures and proximity effects with inhomogeneous order parameters.

Typically, a 38 nm – 100 nm thick layer of $FeF_2$ is grown on a (110) $MgF_2$ substrate at 300ºC, followed by a 4 nm – 70 nm thick ferromagnetic layer (Co, Fe or Ni) grown at 150ºC, coated *in situ* with a 3-4 nm layer of Al to prevent the magnetic layers from oxidation [3]. The $FeF_2$ antiferromagnet ($T_N$=78.4 K) grows epitaxially and untwinned in the (110) orientation on a (110) $MgF_2$ substrate [4] as determined by X-ray diffraction. Based on the bulk structure, the ideal surface of (110) $FeF_2$ is assumed to have compensated spins, oriented in-plane with the easy axis along the [001] direction [3]. X-ray diffraction measurements show that the ferromagnetic layer for all samples is polycrystalline. A uniaxial anisotropy present in all samples, with the easy axis along the $FeF_2$ easy axis, even at temperatures much higher than $T_N$ is attributed to a growth-induced anisotropy. The ferromagnetic transition temperature of all three ferromagnets is well above room temperature. The in-plane sample magnetization parallel to the applied field is measured using a dc SQUID magnetometer and magneto-optical Kerr effect (MOKE). The direction of the sample magnetization during cool down is defined as positive; the sign of the hysteresis loop shift defines the EB sign. The magnetic moment *vs.* applied magnetic field curves for all samples show a typical ferromagnetic behavior above $T_N$: a single hysteresis loop, symmetric with respect to the origin.

In the first series of experiments, the sample is demagnetized at 300 K to a chosen value of the remanent magnetization along the easy axis, $M_R$ between 0 and $M_S$ the saturation magnetization. This leads to the formation of F domains with the magnetizations in opposite directions along the easy axis. The balance between the magnetization of the two types of domains determines the resultant magnetization of the sample. After the sample is cooled in zero magnetic field (ZFC) below $T_N$, the low-field magnetic moment is measured as a function of applied magnetic field at various temperatures. The sample cooled from full remanent magnetization, $M_R \approx M_S$ shows single hysteresis loops exchange biased to negative fields by $H_e(T)$ [Fig. 1(a)]. In contrast, the sample cooled with a reduced remanent magnetization shows double hysteresis loops (e.g. Fig. 1(a) for $M_R \approx 0.5\ M_S$ and $M_R \approx 0$). Each loop is shifted along the magnetic field axis by the same absolute value of temperature-dependent $H_e(T)$, but in the opposite directions. The loop height ratio is set by the remanent state, in which the sample was cooled, and it is equal to the magnetization ratio of the two types of domains in that state. Thus, the system "remembers" the remanent magnetization state.

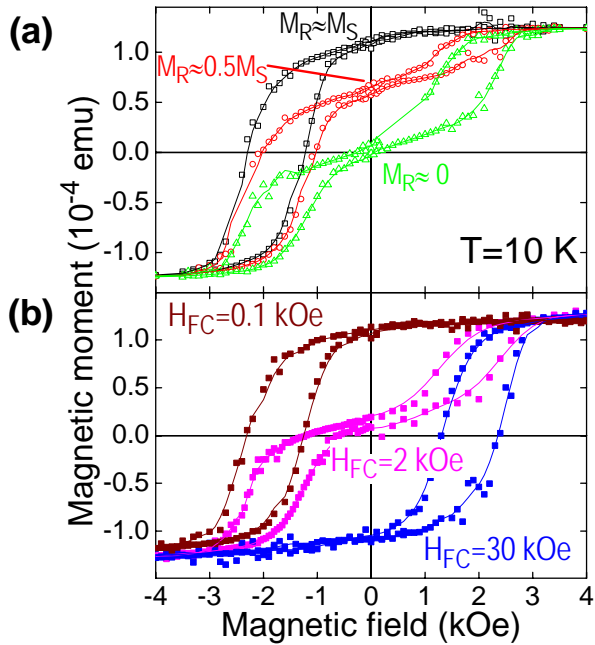

FIG. 1. Easy axis magnetization loops for the FeF$_2$(38 nm)/Co(4 nm) sample below $T_N$, at 10 K, (a) ZFC with three values of the remanent magnetization: $M_S$, $0.5M_S$, 0, and (b) FC in various fields: $H_{FC}$ = 0.1 kOe, 2 kOe, 30 kOe.

In the second series of experiments, the same sample is first magnetized at $T$ = 150 K by applying the magnetic field along the easy axis, above the irreversibility field of the hysteresis curve. Then the sample is cooled in an applied field (FC), $H_{FC}$ [Fig. 1(b)]. For small $H_{FC}$, the magnetization curves below $T_N$ consists of a single hysteresis loop, shifted by negative $H_e(T)$. For large $H_{FC}$, the magnetization curves below $T_N$ also exhibit a single hysteresis loop, but shifted to positive fields by virtually the same absolute value of $H_e(T)$. For intermediate $H_{FC}$, the magnetization curves below $T_N$ consist of two hysteresis loops, one shifted to negative and the other – to the same positive fields $H_e(T)$, as in the first series of experiments. The characteristic cooling field ranges depend on the material and the thickness of both F and AF. For example, for the $FeF_2$(38 nm)/Co(4 nm) sample [Fig. 1(b)], the double loops are observed for cooling fields between 5 kOe and 30 kOe, while for the $FeF_2$(100 nm)/Ni(21 nm) sample the corresponding field range is from 0.5 kOe to 2 kOe.

It is remarkable that in the both experimental series, positive and negative $H_e(T)$ of equal absolute value are found at all temperatures below $T_N$. At any particular temperature, the width of both loops is equal to that of the single hysteresis loop (twice the coercive field, $2H_c$). When rescaled vertically, both the single loops and each of the double loops have exactly the same shape and temperature evolution. Thus, two types of independent regions are formed with identical properties, one positively and the other negatively exchange biased.

It is noteworthy that in these samples the loop half-width is smaller than its shift, *i.e.* $H_c < H_e$. The condition $H_c \leq H_e$ is essential for clear observation of the double hysteresis loops. In contrast, the results presented in [5-6] were in the opposite limit: $H_c \gg H_e$, thus the double loops could not be resolved.

The conclusions of the first two series of experiments are unambiguously confirmed by spatially resolved magneto-optical Kerr effect (MOKE), measured as intensity difference of the reflected *p*-polarized light [7].

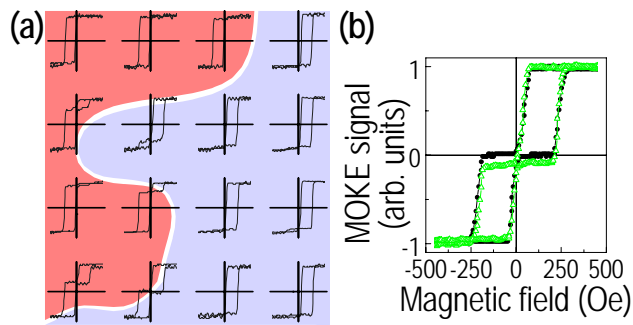

FIG. 2. Magnetic hysteresis loops along the easy axis, measured with MOKE on a 5 mm x 5 mm sample: (a) at different parts of the sample as indicated on the figure, with a ~500 μm laser spot, (b) from the entire sample surface area (black circles), and the average of the 16 curves in (a) (green triangles). The background color in (a) represents the local direction of the EB: red – negative, blue – positive.

For this experiment, a 5 mm × 5 mm $FeF_2$(70 nm)/Ni(70 nm)/Al(4 nm) sample is demagnetized above $T_N$ and then cooled below $T_N$ in a zero applied magnetic field as it is done in the first series of the experiments. First, the MOKE signal (generally considered [7] to be proportional to the magnetization) as a function of applied magnetic field is collected from the entire sample surface area, illuminated with a wide beam. The curve consists of a double hysteresis loop [Fig. 2(b)]. Without any change in the experimental conditions, MOKE measurements are performed using a ~500 μm diameter laser beam at 16 spots arranged in a 4×4 matrix, in the positions shown in the Fig. 2(a). The resultant signal [Fig. 2(a)], which is proportional to the magnetization, varies spatially: on one side of the sample the single loop is negatively shifted, on the other side – the single loop is positively shifted, and in between – double hysteresis loops are found. The normalized sum of these 16 curves [green triangles in Fig. 2(b)] is in good agreement with the hysteresis curve obtained from the entire sample (black circles). The slight difference between the two curves is due to incomplete coverage of the surface area by the 16 measured spots.

This experiment confirms that the sample has two types of areas; single hysteresis loops are shifted by the same $H_e$, positively in one area, and negatively in the other. When the laser beam covers parts of both areas, the magnetization curve consists of two loops shifted in opposite directions, with their relative heights determined by the ratio of the two areas. The magnetization from the whole sample corresponds to the sum of magnetization from all its individual parts.

These experimental results lead to two main findings. *First*, the sample can split into two independent subsystems with opposite sign of the EB. This can be achieved either by cooling a partially or completely demagnetized sample in zero applied magnetic field, or by cooling the sample in a judiciously chosen intermediate applied magnetic field. The appropriate field range is determined by sample parameters: materials, layer thickness, interface roughness, etc. *Second*, the exchange bias for each F domain is not averaged over various AF domains, so double hysteresis loops are observed.

The phenomenon of EB [8] has attracted much attention [9-11]; several mechanisms of the EB have been proposed [8-13] of which many neglect variations of EB on the scale of F domains. The dependence of $H_c$ and $H_e$ on the AF and/or F domain sizes in patterned F-AF bilayers [14], or in diluted AF [15] has been studied. Enhancement of $H_e$ in the AF-F bilayers with patterned F layer was explained by suppression of F-F exchange interaction in the small F domains [16]. Correspondence of

AF domains to the F domains [5], and small spatial variations of the value of EB [6,17] were reported. Double hysteresis loops [18] and a correlation of domain structure [19] in the F and the AF were observed in CoO/NiFe samples zero-field cooled from a demagnetized state and explained using the exchange spring model [18,19]. Exchange bias field and the shape of the double hysteresis loops in $FeF_2$/Fe [20] and $Fe_{0.6}Zn_{0.4}F_2$/Fe [21] bilayers were found to depend on the remanent magnetization during cooling in a zero applied magnetic field. Double and negatively shifted single hysteresis loops were observed in FeMn/FeMnC samples [22] deposited in various small magnetic fields (0-40 Oe) without any clear trend. In zero-field deposited NiFe/NiFeMn samples, spatial variation of MOKE signal measured along the hard axis was claimed [23], but no supporting plot was presented.

The large number of reported experiments performed on different samples, show seemingly contradictory and unconnected results. Combined with multiple interpretations, this leads to a complex and confusing situation where no conclusion can be made. It is obvious that only a comprehensive study with several different experiments on the same sample can provide definite answers on the role of the relative sizes of the AF and F domains in EB.

In EB, we find two distinct regimes depending on the relation between the AF and F domain sizes. First, when the AF domains are smaller than the F domains, a F domain averages the direction and magnitude of $H_e$ over several AF domains. This may lead to a continuous variation of $H_e$ as a function of the cooling field and single hysteresis loops, as found, in twinned $FeF_2$/Fe [24] and $MnF_2$/Fe [25] samples. If the AF domains are very small, the number of the AF domains interacting with each F domain is very large. In this case, each F domain can be essentially treated as a separate sample. When ZFC the sign for the net EB for each of these domains is set by the direction of the magnetization of the F domain, so double loops may occasionally be observed [20]. X-ray reflectivity measurements [4] imply that the in-plane structural coherence in the twinned (110) $FeF_2$ grown on (100) MgO is 6-10 nm with the easy axes of the crystallites at 90° to each other. Such spatial variations of the easy axis, combined with the small size of the AF domains, may result in the averaging regime of EB.

When the AF domain size is comparable to or larger than the F domain size, each F domain couples only to one AF domain with a particular direction of the EB, so no averaging occurs. This results in the same absolute value of the $H_e$ for different cooling fields and double hysteresis loops, as reported in this work. In this regime, when the sample is zero-field cooled, the unidirectional anisotropy in the AF is set locally through the antiferromagnetic coupling [24] between the interfacial F and AF moments. Therefore, the local sign of the EB for each F domain is determined by the direction of the magnetization of that domain during sample cooling. Thus, domains of the F with opposite magnetization direction above $T_N$ have the same magnitude but opposite sign of EB below $T_N$.

When the sample is field cooled, also, only the sign, but not the magnitude of $H_e$ depends on the cooling field, $\boldsymbol{H}_{FC}$. It is the sign of the projection of $\boldsymbol{H}_{total}$ on the easy axis, $H_{total}$, that determines the sign of the local EB. $\boldsymbol{H}_{total}$, the total local cooling field sensed by the interfacial spins of the AF, is the sum of the applied magnetic field, $\boldsymbol{H}_{FC}$, and the local exchange field due to the interfacial F moments, $\boldsymbol{H}_F$. Spatial inhomogeneity of the sample leads to inhomogeneity of $\boldsymbol{H}_F$, which in turn leads to inhomogeneity of $H_{total}$. At intermediate $H_{FC}$, that causes the magnitude of $H_{total}$ to be small and spatially inhomogeneous, giving rise to areas of the interface with positive and negative sign of $H_{total}$. Consequently, cooling the sample below $T_N$ in this field leads to the negative and the positive EB, respectively. According to X-ray reflectivity measurements [4] the in-plane structural coherence in the untwinned (110) $FeF_2$ grown on a (110) $MgF_2$ single crystal substrate is ~ 28 nm. Moreover, due to absence of twinning, the size of AF domains can be much larger than the grain size, unlike in the case of a twinned $FeF_2$.

The relevance of the observation presented here to other systems was illustrated in a multilayer system imitating an AF-F system [26]. For the sample with the larger domains in the layer playing the role of interfacial AF spins, double hysteresis loops were observed for intermediate cooling fields. It is not clear whether such a model system correctly mimics the domain structure in real AF-F systems. Those results, however, provide an additional experimental configuration in which relative domain sizes play a role.

In conclusion, we report for the first time that the relative AF and F domain sizes affect exchange bias in a fundamental fashion. When the AF domain size is larger than or comparable to that in the F, structural or magnetic inhomogeneities can result in the sample splitting into two independent subsystems upon cooling through the AF transition temperature, $T_N$. Each subsystem exhibits EB of the same magnitude but of the opposite sign. This results in magnetization curves with double-hysteresis loops, thus unambiguously showing non-averaging, local EB. This behavior is reproducibly observed in samples with three different F: Ni, Fe, and Co, with various F and AF layer thicknesses. These observations also show that not only the properties of the F but those of the AF are modified in EB.

We suggest that relative lateral length scales maybe relevant in other proximity effects with spatially inhomogeneous order parameters, such as a granular superconductor (S) in proximity with a normal metal, or a S in proximity with a F that has domains. Recently, a theoretical prediction [27] and an experimental observation [28] were reported for a S-F proximity system where the properties of the type-II S, including its transition temperature, $T_c$, were modified depending on the presence or absence of domains in the F. It is very likely that the relation between the size of the domains or the domain wall width in the F and the coherence length or magnetic screening length in the S could drastically affect the proximity effect in S-F systems.

The work is funded by the US Department of Energy and the AFOSR. Also, financial support from Alexander-von-Humboldt Foundation (O. P.), Cal(IT)$^2$ (Z.-P. L.), Spanish MECD (R. M., X. B.), Fulbright Commission (R. M.), and Catalan DURSI (X. B.) is acknowledged. We thank M. R. Fitzsimmons, E. E. Fullerton, A. Hoffmann, M. Kiwi, D. Lederman, C. Leighton, K. Liu, Yu. Lyanda-Geller, C. Miller, S. Roy, S. K. Sinha, and H. Suhl for their valuable comments, and J. M. Alameda for his help with the magneto-optical measurements.

[1] For EB treated as a proximity effect see e.g. P. J. van der Zaag *et al.*, *Phys. Rev. Lett.* **84**, 6102 (2000), H. Suhl, I. K. Schuller, *Phys Rev. B* **58**, 258 (1998).

[2] See e.g. H. Ohldag *et al.*, *Phys. Rev. Lett.* **86**, 2878 (2001).

[3] J. Nogués *et al.*, *Phys. Rev. B* **59**, 6984 (1999).

[4] H. Shi, D. Lederman, *Phys. Rev. B* **66**, 094426 (2002).

[5] P. Kappenberger *et al.*, *Phys. Rev. Lett.* **91**, 267202 (2003).

[6] F. Nolting *et al.*, Nature **405**, 767 (2000).

[7] P. M. Oppeneer, in *Handbook of Magnetic Materials*, edited by K. H. J. Buschow, (Elsevier, Amsterdam 2001) Vol. 13, p. 229.

[8] W. H. Meiklejohn, C. P. Bean, *Phys. Rev.* **102**, 1413 (1956).

[9] J. Nogués, I. K. Schuller, *J. Magn. Magn. Mater.* **192**, 203 (1999).

[10] A. E. Berkowitz, K. Takano, *J. Magn. Magn. Mater.* 200, 552 (1999).

[11] R. L. Stamps, *J. Phys. D: Appl. Phys.* **33**, R247 (2000).

[12]. M. Kiwi, *J. Magn. Magn. Mater.* **234**, 584 (2001).

[13] A. P. Malozemoff, *Phys. Rev. B* **35**, 3679 (1987).

[14] M. Fraune *et al.*, *Appl. Phys. Lett.* **77**, 3815 (2000).

[15] P. Miltényi *et al.*, *Phys. Rev. Lett.* **84**, 4224 (2000).

[16] K. Liu *et al.*, *Phys. Rev. B* **63**, 060403 (2001).

[17] F. Offi *et al.*, *J. Magn. Magn. Mater.* **261**, L1 (2003).

[18] N. J. Gökemeijer, J. W. Cai, C. L. Chien, *Phys. Rev. B* **60**, 3033 (1999).

[19] C. L. Chien *et al.*, *Phys. Rev. B* **68** 014418 (2003).

[20] P. Miltényi *et al.*, *Appl. Phys. Lett.* **75**, 2304 (1999).

[21] Ch. Binek *et al.*, *J. Magn. Magn. Mater.* **240**, 257 (2002).

[22] H.-W Zhao *et al.*, *J. Appl. Phys.* **91**, 6893 (2002).

[23] C.-H. Lai, S.-A. Chen, A. Huang, *J. Magn. Magn. Mater.* **209**, 122 (2000).

[24] J. Nogués *et al.*, *Phys Rev. Lett.* **76**, 4624 (1996).

[25] C. Leighton *et al.*, *Phys. Rev. Lett.* **84**, 3466 (2000).

[26] T. L. Kirk, O. Hellwig, E. E. Fullerton, *Phys. Rev. B* **65**, 224426 (2002).

[27] A.Yu. Aladyshkin *et al.*, *Phys. Rev. B* **68**, 184508 (2003).

[28] M. Lange, M. J. Van Bael, V. V. Moshchalkov, *Phys Rev. B* **68**, 174522 (2003).